\documentclass[twocolumn,preprintnumbers,amsmath,amssymb,showpacs]{revtex4}

\usepackage{graphicx}
\usepackage{dcolumn}
\usepackage{bm}

\begin{document}

\title{Effective Temperatures in Athermal Systems Sheared at Fixed Normal Load}

\author{Ning Xu$^1$}
\author{Corey S. O'Hern$^{1,2}$}
\affiliation{$^1$~Department of Mechanical Engineering, Yale University, 
New Haven, CT 06520-8284.\\
$^2$~Department of Physics, Yale University, New Haven, CT 06520-8120.\\
}
\date{\today}

\begin{abstract}

We perform molecular dynamics simulations of repulsive athermal
systems sheared at fixed normal load to study the effective
temperature $T_L$ defined from time-dependent fluctuation-dissipation
relations for density.  We show that these systems possess two
distinct regimes as a function of the ratio $T_S/V$ of the granular
temperature to the potential energy per particle.  At small $T_S/V$,
these systems are pressure-controlled and $T_L$ is set by the normal
load.  In contrast, they behave as quasi-equilibrium systems with $T_L
\approx T_S$ that increases with shear rate at large $T_S/V$.  These results
point out several problems with using $T_L$ in thermodynamic
descriptions of slowly sheared athermal systems.
\end{abstract}

\maketitle

Athermal and glassy systems driven by shear or other external forces
are by definition not in thermal equilibrium.  Fluctuations in these
systems are induced by the driving force and do not arise from random
thermal motion.  However, there are obvious similarities between
equilibrium thermal systems and driven systems that reach a
nonequilibrium steady-state.  In fact, there have been many attempts
to develop thermodynamic and statistical descriptions \cite{jou} of
these systems including kinetic theories for driven granular gases
\cite{jenkins}, the Edward's entropy formalism for granular packings
\cite{edwards,coniglio}, and applications of equilibrium linear
response and fluctuation-dissipation (FD) relations \cite{cugliandolo}
to define effective temperatures in aging
\cite{barrat_glass,dileonardo} and sheared glasses
\cite{sheared_liquid,ohern_puzzle} and compacting
\cite{colizza,coniglio2} and sheared granular materials
\cite{makse,kondic}.

The approach that employs FD relations to define effective
temperatures in athermal and glassy systems has shown great promise.
First, effective temperatures from FD relations for several quantities
such as density and pressure have been shown to be the same
\cite{sheared_liquid,ono}. Second, FD relations can be applied to
dense systems with elastic particles in contrast to other approaches.
FD relations can also be measured experimentally in athermal systems
such as foams and granular materials.  However, it is still not clear
whether effective temperatures from FD relations can be used in
thermodynamic descriptions of dense shear flows.  Many important
questions remain unanswered, for example, what variables should be
used to construct an equation of state and do effective temperature
gradients generate heat flow?  We conduct molecular dynamics (MD)
simulations of sheared, athermal systems to begin to address these
questions.

Nearly all numerical simulations that have measured FD relations in
sheared athermal and glassy systems have been performed at constant
volume.  We will conduct our simulations at constant normal load (or
external pressure), instead, and this has several advantages.  First,
it has been shown that sheared, athermal systems behave differently in
these two ensembles.  At constant normal load, these systems can
dilate in response to an applied shear stress
\cite{nedderman,thompson}.  We will determine how the ability to
dilate affects FD relations and properties of effective temperatures
defined from them.  Simulations at constant normal load will also
enable us to determine whether the effective temperature is more
sensitive to changes in pressure or shear stress.

Our principal result is that the effective temperature $T_L$ defined
from FD relations for density is controlled by pressure in
slowly-sheared athermal systems.  We find that for all shear rates
below ${\dot \gamma}_c$, the effective temperature is independent of
shear rate at fixed normal load. Above ${\dot \gamma}_c$, $T_L$
increases with shear rate and as ${\dot \gamma}$ increases further the
systems obey quasi-equilibrium FD relations at all times.  The
characteristic shear rate that separates the pressure-controlled from
the quasi-equilibrium regime can be estimated using the ratio of
kinetic to potential energy \cite{kondic}.  When the ratio is small, a
strong force network exists and the effective temperature is set by
the pressure.  When kinetic energy dominates, velocity fluctuations are
large, which causes frequent rearrangements, diffusion, and
quasi-equilibrium behavior.

We now provide the essential details of the MD simulations. The
systems contained $N/2$ large and $N/2$ small particles with equal
mass $m$ and diameter ratio $1.4$ to prevent shear induced
ordering and segregation \cite{onuki}.  The particles
interacted via one of the following pairwise, purely repulsive
potentials:
\begin{eqnarray} 
\label{spring} 
V^S(r_{ij}) & = & \frac{\epsilon}{\alpha} 
\left(1-r_{ij}/\sigma_{ij}\right)^{\alpha}  \\ 
\label{rlj}
V^{RLJ}(r_{ij}) & = & \frac{\epsilon}{72} 
\left[({\sigma_{ij}/r_{ij}})^{12} - 2({\sigma_{ij}/r_{ij}})^6 + 1\right],
\end{eqnarray} 
where $\epsilon$ is the characteristic energy scale and
$\sigma_{ij}=\left(\sigma_i + \sigma_j \right) /2$ and $r_{ij}$ are
the average diameter and separation of particles $i$ and $j$.  The
repulsive linear ($\alpha=2$) and Hertzian ($\alpha=5/2$) spring
potentials (Eq.~\ref{spring}) have been used to model granular
materials \cite{luding,silbert}.  In contrast to Eq.~\ref{spring}, the
repulsive Lennard-Jones (RLJ) potential (Eq.~\ref{rlj}) has an
infinite repulsive core at small $r_{ij}$.  Both interaction
potentials are zero for $r_{ij}\ge\sigma_{ij}$.

For athermal dynamics, the position and velocity of each 
particle in the bulk is obtained by solving
\begin{equation}
m\frac{d^2\vec{r_i}}{dt^2}=\vec{F}^r_i-b\sum_j\left( \vec{v}_i-\vec{v}_j
\right),
\end{equation}
where $\vec{F}^r_i=-\sum_j dV(r_{ij})/dr_{ij} \hat{r}_{ij}$ is the
total repulsive force on particle $i$, $\vec{v}_i$ is the velocity of
particle $i$, $b>0$ is the damping constant, and sums over $j$ only
include particles overlapping particle $i$.  Shear is imposed by
moving a rough and disordered top boundary in the $x$-direction at
fixed speed $u$, while a similar bottom wall a distance $L_y$ away
remains stationary.

To sustain a fixed normal load $P_{ext}$ in the $y$-direction, the top
boundary was moved rigidly according to
\begin{equation}
M\frac{d^2L_y}{dt^2}=F^w_y-F_{ext}-b\sum_j\left( \frac{dL_y}{dt}-v_{yj}
\right),
\end{equation}
where $M$ is the mass of the top wall, $F^w_y$ is the total repulsive
force acting on the top wall from interactions with particles in the
bulk, and $F_{ext} = P_{ext} L_x L_z$ ($P_{ext}L_x$) is the external
normal force applied to the top wall in 3d (2d).  The equations of
motion were solved using a $5$th order Gear predictor-corrector
integration scheme \cite{allen}.  After an initial transient at each
$u$ and $P_{ext}$, forces in the $y$-direction balance on average, the
system fluctuates about $\langle L_y \rangle$, and a linear shear flow
is established with shear rate ${\dot \gamma} = u/\langle L_y
\rangle$.  Relatively small system sizes $N=1024$ were studied to
inhibit the formation of nonlinear velocity profiles \cite{ning}.
Periodic boundary conditions were implemented in the $x$ ($z$)
direction.  We use $\sigma$, $\epsilon$, and
$\sigma/\sqrt{m/\epsilon}$, where $\sigma$ is the small particle
diameter, as the units of length, energy, and time, respectively.  All
quantities are expressed in reduced units below.

\begin{figure}
\scalebox{0.4}{\includegraphics{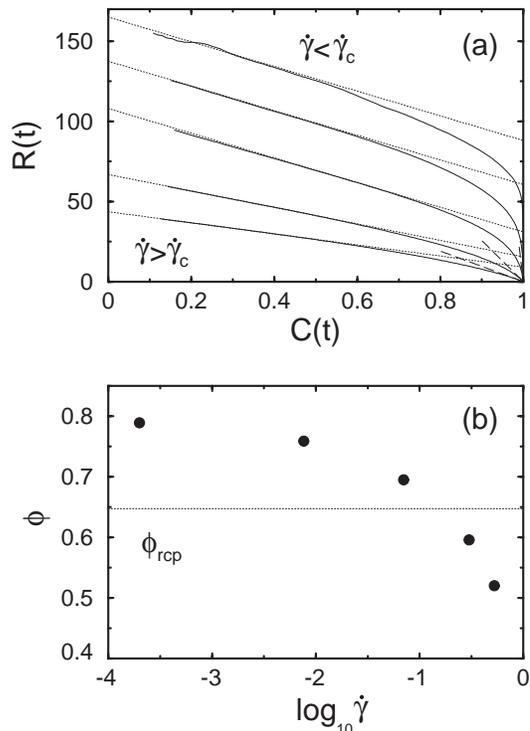}}%
\vspace{-0.18in}
\caption{\label{fig:fig1} (a) $R(t)$ vs. $C(t)$ (solid lines) for
density $\rho({\vec k})$ at ${\vec k} = 9 {\hat z}$ for 3d systems
with RLJ interactions and $b=0.5$ sheared at fixed normal load
$P_{ext}=0.1$.  The dotted (dashed) lines have slope $-1/T_L$
($-1/T_S$).  Five shear rates are shown: ${\dot \gamma}= 0.53$,
$0.30$, $0.07$, $7.7 \times 10^{-3}$, and $2.0 \times 10^{-4}$ from
bottom to top.  $T_L$ is constant for ${\dot \gamma} < {\dot \gamma}_c
\approx 0.2$. (b) Volume fraction $\phi$ vs. ${\dot \gamma}$ for the
same systems in (a). $\phi_{rcp} \approx 0.648$ for 3d bidisperse systems.}
\vspace{-0.22in}
\end{figure}

In equilibrium systems, the fluctuation-dissipation theorem requires
that the autocorrelation function of a physical quantity is
proportional at {\it all} times $t$ to that quantity's response to a
small conjugate perturbation, and the proportionality constant gives
the temperature.  Response and correlation are not proportional at all
times in sheared glassy and athermal systems
\cite{cugliandolo}. However, several recent studies of FD relations
for density in these systems have shown that they can still be used to
define an effective temperature that characterizes shear-induced
fluctuations at long-time scales
\cite{sheared_liquid,ohern_puzzle}. Similar measurements in 3d sheared
athermal systems, but at fixed normal load, are displayed in
Fig.~\ref{fig:fig1} (a).  We plot the integrated response $R(t)$ of
the density of the large particles at wavevector ${\vec k} = k_0 {\hat
z}$
\begin{equation}
\label{density}
\rho(\vec{k},t)=\frac{1}{N_L} \sum^{N_L}_{j=1}e^{i\vec{k}\cdot \vec{r}_j(t)}
\end{equation} 
to a spatially modulated force with period $2\pi/k_0$ applied along
${\vec k}$ at $t=0$ and all subsequent times versus the density
autocorrelation function $C(t)$ \cite{remark}.  This figure shows that
response versus correlation for $\rho({\vec k})$ can be nonlinear at
short times in sheared athermal systems.  In contrast, the slope of $R(C)$ is
constant at long times, which allows the long-time effective
temperature $T_L$ to be defined as
\begin{equation}
-1/T_L = \left. \frac{dR}{dC} \right|_{C\rightarrow 0}.
\end{equation}
Previous studies have shown that $T_L$ does not depend strongly on the
magnitude \cite{sheared_liquid} or direction of ${\vec k}$
\cite{ohern_puzzle}.  The slope of $R(C)$ for $\rho({\vec k})$ at
${\vec k} = k_0 {\hat z}$ in the $t \rightarrow 0$ limit is $-1/T_S$, where
$T_S = m \langle v_z^2\rangle$ is the granular temperature.

\begin{figure}
\scalebox{0.4}{\includegraphics{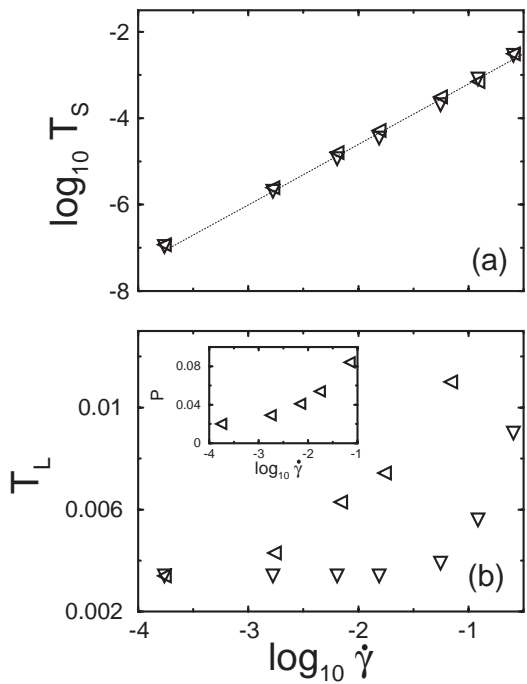}}%
\vspace{-0.18in}
\caption{\label{fig:fig4} (a) $T_S$ and (b) $T_L$ vs. ${\dot \gamma}$
for 3d systems with RLJ interactions and $b=0.5$ sheared at fixed
$P_{ext}=0.02$ (downward triangles) or fixed volume fraction
$\phi=0.68$ (leftward triangles).  The dotted line in (a) has slope
$1.4$. The inset to (b) shows the internal pressure $P$ vs. ${\dot \gamma}$ at
fixed $\phi=0.68$.}
\vspace{-0.22in}
\end{figure}

In Fig.~\ref{fig:fig1} (a), we show $R(C)$ for several shear rates at
fixed normal load $P_{ext}=0.1$.  At low shear rates, the long-time
slope of $R(C)$ is independent of ${\dot \gamma}$.  However, when ${\dot
\gamma}$ increases above a characteristic shear rate ${\dot
\gamma}_c$, the long-time slope of $R(C)$ begins to decrease with
increasing ${\dot \gamma}$.  Thus, at constant normal load, $T_L$ is
constant for ${\dot \gamma} < {\dot \gamma}_c$, but increases for
${\dot \gamma} > {\dot \gamma}_c$.  Fig.~\ref{fig:fig1} (b) shows the
variation of the volume fraction $\phi$ with ${\dot \gamma}$ for
the systems in (a).  To compensate for increases in normal force from
increases in ${\dot \gamma}$, the systems expand in the $y$-direction.
Fig.~\ref{fig:fig1} (b) emphasizes that effective temperatures depend
on the time scale over which they are measured even in systems that
are $20\%$ below random close packing.

The results for $R(C)$ for $\rho({\vec k})$ at constant normal load
presented in Fig.~\ref{fig:fig1}(a) are intriguing; they suggest that
$T_L$ is fixed by the pressure in slowly sheared systems.  To
investigate this question further, we compare $T_S$ and $T_L$ in
systems sheared at constant normal load (CN) and constant volume
fraction (CV) in Fig.~\ref{fig:fig4}.  We find that $T_S$ is
insensitive to the choice of the ensemble; $T_S$ scales as a power-law
in ${\dot \gamma}$ over three decades in both ensembles.  However, the
${\dot \gamma}$ dependence of $T_L$ at CN and CV differs
significantly.  $T_L$ in the two ensembles were initially matched at
${\dot \gamma} \approx 10^{-4}$ and $P_{ext}=0.02$ by setting $\phi$
in the CV simulation equal to $\langle \phi \rangle$ from the
simulation at fixed $P_{ext}=0.02$.  At fixed $\phi$, $T_L$ increases
by more than a factor of $3$ over the range of ${\dot \gamma}$
studied.  The inset to Fig.~\ref{fig:fig4} (b) suggests that the
steady rise in $T_L$ is due to the shear-induced increase in the
internal pressure $P$ at CV.  In contrast, $T_L$ at fixed $P_{ext}=0.02$
remains {\it constant} over the same range of ${\dot \gamma}$.  $T_L$
only begins to increase when ${\dot \gamma} > {\dot \gamma}_c$, which
in this case is at least $3$ decades above the quasi-static ${\dot
\gamma}$ regime at CV.

\begin{figure}
\scalebox{0.48}{\includegraphics{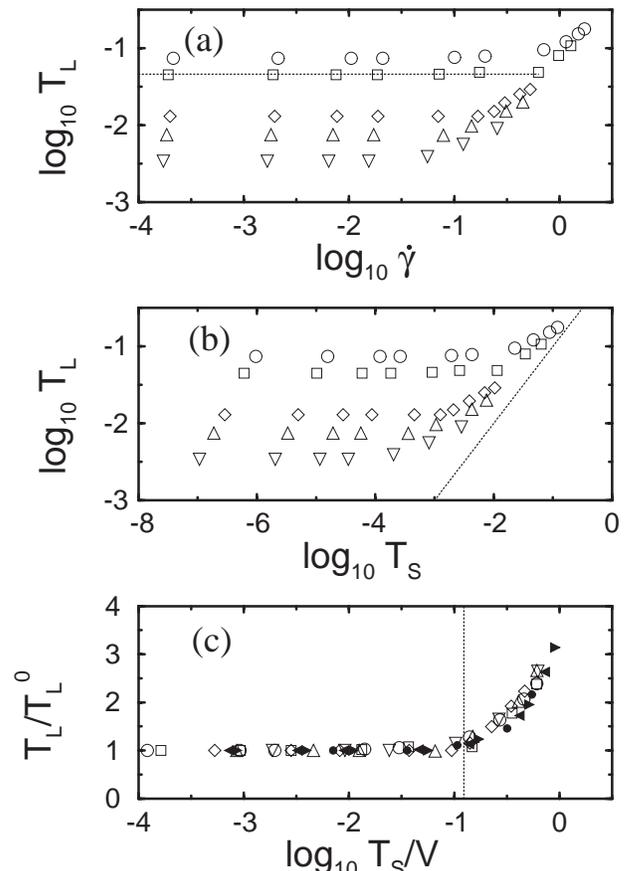}}%
\vspace{-0.2in}
\caption{\label{fig:fig2} $T_{L}$ plotted vs. (a) ${\dot \gamma}$ and
(b) $T_S$.  Panel (c) shows $T_L/T_L^0$ vs. the ratio of $T_S$ to the
total potential energy per particle $V$.  The results were obtained
from 3d systems with RLJ interactions at fixed $P_{ext}$.  Five loads
$P_{ext}=0.02$ (downward triangles), $0.05$ (upward triangles), $0.1$
(diamonds), $0.5$ (squares), and $1$ (circles) and three loads
$P_{ext} = 2$ (rightward triangles), $5$ (leftward triangles), and
$10$ (small circles) were studied at $b=0.5$ (open symbols) and $0.04$ (filled
symbols), respectively.  In panel (a), the dotted line defines the low
shear rate value $T_L^0$ of the long-time effective temperature at
$P_{ext}=0.5$. In panel (b), $T_L=T_S$ is indicated by the dotted
line.  In panel (c), the dotted line marks the ratio of kinetic to
potential energy above which $T_L/T_L^0$ begins to increase for these
systems.}
\vspace{-0.22in}
\end{figure}

To test the generality of these results, we performed measurements of
$R(C)$ for $\rho({\vec k})$ at two values of the
damping constant ($b=0.5$ and $0.04$), over a range of normal loads
from $P_{ext}=0.02$ to $10$, and over $4$ decades in shear rate
\cite{remark2}. The results are summarized in Fig.~\ref{fig:fig2}.
Panel (a) shows that for each $P_{ext}$, there is a wide range of
${\dot \gamma}$ over which $T_L=T_L^0$ is constant.  $T_L$ begins to
increase above a characteristic shear rate ${\dot \gamma}_c$, which
decreases with normal load.  Fig.~\ref{fig:fig2} (b) shows that $T_L$
begins to deviate from $T_L^0$ as $T_S$ increases.  In this regime,
large velocity fluctuations give rise to frequent rearrangement events
and particle diffusion.  At each $P_{ext}$, there is a $T_S$ (or
${\dot \gamma}$) regime where sheared athermal systems behave as
quasi-equilibrium systems with response proportional to correlation at
all times and a single effective temperature $T_S \approx T_L$ that
increases with shear rate.  This nontrivial result is not found in
sheared glasses because they are thermostatted with $T_S$ below the
glass transition temperature and remain arbitrarily far from
equilibrium as the shear rate is tuned.

We seek a general criterion to determine whether sheared athermal
systems exist in the pressure-controlled or quasi-equilibrium regime.
As a first step, we consider the ratio $\beta$ of the granular
temperature $T_S$ to the total potential energy per particle $V$.  For
each value of $b$, we find that there is a range of normal loads
$P_{ext} > P_{ext}^c$ where $T_L/T_L^0$ versus $T_S/V$ is independent
of $P_{ext}$, with $P_{ext}^c$ increasing inversely with $b$.
Fig.~\ref{fig:fig2} (c) shows that when $P_{ext}>P_{ext}^c$, there is
a characteristic ratio $\beta_c \approx 0.1$ independent of damping
that separates the pressure-controlled and quasi-equilibrium regimes.
These results also imply that $T_L$ in sheared glasses is controlled
by pressure since they have $T_S/V \ll 1$.  The behavior at large
$T_S/V$ in the low-pressure regime $P_{ext}<P_{ext}^c$ is more complex
and will be discussed elsewhere \cite{ning2}.

\begin{figure}
\scalebox{0.4}{\includegraphics{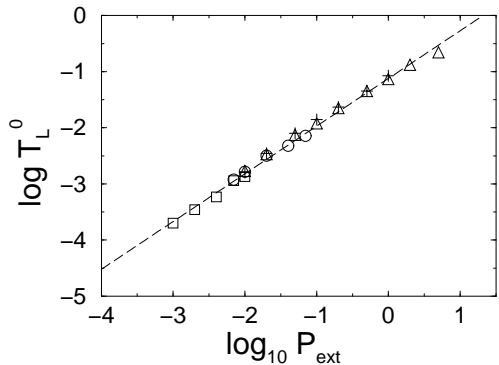}}%
\vspace{-0.2in}
\caption{\label{fig:fig3} $T_L^0$ vs. $P_{ext}$ for several sheared
athermal systems at $b=0.5$: 3d RLJ (triangles), 2d RLJ (pluses), 3d
linear spring (circles), and 3d Hertzian spring (squares). The dashed
line has slope $0.85$.  $P_{ext} > 10^{-1}$ ($10^{-2}$) were not
considered for linear (Hertzian) springs because they caused
unphysical overlaps.}
\vspace{-0.2in}
\end{figure}

In Fig.~\ref{fig:fig3} we plot the long-time effective temperature
$T_L^0$ in the pressure-controlled regime vs. $P_{ext}$ for sheared
athermal systems in 2d and 3d with repulsive linear and Hertzian
spring and Lennard-Jones interactions.  All of the data collapse onto
a power-law with exponent $0.85 \pm 0.02$.  If both the normal load
\cite{ghe} and $T_L^0$ scaled linearly with the yield stress of the
material, the exponent in Fig.~\ref{fig:fig3} would be $1$.  Our
preliminary results indicate that the yield stress and pressure are
proportional \cite{ning2}, which implies that $T_L^0$ scales
sublinearly with the yield stress.

We used MD simulations to study properties of the effective
temperature $T_L$ defined from FD relations for density in athermal
systems sheared at fixed normal load.  These systems possess two
distinct regimes as a function of the ratio $T_S/V$ of kinetic to
potential energy.  At small ratios, these systems are
pressure-controlled and $T_L$ is set by the normal load.  At large
$T_S/V$, they behave as quasi-equilibrium systems with $T_L \approx
T_S$ that increases with shear rate.  These results point out several
difficulties in using $T_L$ in thermodynamic descriptions of slowly
sheared athermal systems.  First, the variables $T_L$, pressure, and
density alone do not provide a complete description of these systems
since $T_L$ and pressure can remain constant while density can vary
substantially.  Also, these results suggest that when two slowly
sheared, athermal systems are placed in contact but maintained at
different pressures, $T_L$ in the two systems will not equilibrate.
Thus, in this regime $T_L$ does not behave as a true temperature
variable.  We are currently attempting to identify a set of variables
that can be used in a thermodynamic description of dense
granular shear flows.

We thank R. Behringer, L. Kondic, and A. Liu for helpful
comments. Financial support from NASA grant NAG3-2377 (NX) and Yale
University (NX,CSO) is gratefully acknowledged.


\begin{references}
\bibitem{remark} Large particles that were less than one large
particle diameter from the boundaries were not included in
Eq.~\ref{density} to avoid boundary effects; thus, $N_L < N/2$.
\bibitem{remark2} At each ${\dot \gamma}$ and $P_{ext}$, we verified that
the response was in the linear regime and averaged over at least $500$
independent configurations.  We also calculated the response and
correlation functions at two system sizes, $N=256$ and $1024$, to show
that finite-size effects were small.
\end{references}
\end{document}